\documentclass{ws-ijmpd}
\usepackage[super,compress]{cite}

\begin{document}

\markboth{Dirk Puetzfeld and Yuri N.\ Obukhov}
{Prospects of detecting spacetime torsion}

%
\catchline{}{}{}{}{}
%

\title{PROSPECTS OF DETECTING SPACETIME TORSION\footnote{This essay received a honorable mention in the 2014 essay competition of the Gravity Research Foundation.}}

\author{DIRK PUETZFELD}

\address{ZARM, University of Bremen\\
        Am Fallturm, 28359 Bremen, Germany\\
dirk.puetzfeld@zarm.uni-bremen.de, http://puetzfeld.org}

\author{YURI N.\ OBUKHOV}

\address{Theoretical Physics Laboratory, Nuclear Safety Institute, Russian Academy of Sciences\\ 
 B.Tulskaya 52, 115191 Moscow, Russia\\
obukhov@ibrae.ac.ru}

\maketitle

\begin{history}
\received{Day Month Year}
\revised{Day Month Year}
\end{history}

\begin{abstract}
How to detect spacetime torsion? In this essay we provide the theoretical basis for an answer to this question. Multipolar equations of motion for a very general class of gravitational theories with nonminimal coupling in spacetimes admitting torsion are given. Our findings provide a framework for the systematic testing of whole classes of theories with the help of extended test bodies. One surprising feature of nonminimal theories turns out to be their potential sensitivity to torsion of spacetime even in experiments with ordinary (not microstructured) test matter.
\end{abstract}


\ccode{PACS numbers: 04.50.Kd; 04.20.Cv; 04.20.Fy}


\section{Introduction} \label{sec_introduction}

The dynamics of test bodies in curved manifolds represents an interesting problem of determining the geometrical structure of spacetime. In Einstein's General Relativity (GR) spacetime is a Riemannian manifold. However, this is not necessarily true in gauge gravitational theories, which can be considered as viable alternatives to GR. 

As Einstein \cite{Einstein:1921} himself formulated:
\begin{quote}
``[...] The question whether this continuum has a Euclidean, Riemannian, or any 
other structure is a question of physics proper which must be answered by 
experience, and not a question of a convention to be chosen on grounds
of mere expediency.''
\end{quote}

One of the possible non-Riemannian deviations of the geometry of spacetime is torsion (introduced very early by E.\ Cartan). The physical problem is then as follows: can we detect the torsion of spacetime and how we can do it? 

Torsion arises on an equal footing with the curvature as the gravitational field strength in the gauge-theoretic approach to gravity \cite{Hehl:1995,Blagojevic:Hehl:2013}. The sources of these Poincar\'e gauge fields are the two Noether currents: the mass (energy-momentum) and spin. Accordingly, an early analysis of the problem of motion revealed that curvature and torsion can be detected by means of test bodies composed of microstructed matter \cite{Stoeger:Yasskin:1980}, the elements of which have mass and spin. In agreement with this analysis, in recent experimental efforts \cite{Hunter:etal:2013,Lehnert:etal:2014} one uses polarized test bodies for placing limits on the torsion of spacetime. 

In this essay we demonstrate a possibility of detecting the effects of torsion due to the nonminimal coupling of matter with the gravitational field. 

\section{General nonminimal gravity} \label{sec_gennonmin}

As in Ref.~\refcite{Stoeger:Yasskin:1980}, we consider matter with microstructure, namely, with spin. An appropriate gravitational model is then the Poincar\'e gauge theory in which the metric tensor $g_{ij}$ is accompanied by the connection $\Gamma_{ki}{}^j$ that is metric-compatible but not necessarily symmetric; for details see Refs.~\citen{Hehl:1995,Blagojevic:Hehl:2013}. The gravitational field strengths are the Riemann-Cartan curvature and the torsion:
\begin{eqnarray}
R_{kli}{}^j &=& \partial_k\Gamma_{li}{}^j - \partial_l\Gamma_{ki}{}^j + \Gamma_{kn}{}^j \Gamma_{li}{}^n - \Gamma_{ln}{}^j\Gamma_{ki}{}^n,\label{curv}\\
T_{kl}{}^i &=& \Gamma_{kl}{}^i - \Gamma_{lk}{}^i.\label{tors}
\end{eqnarray}

We consider a general nonminimal gravity model in which the interaction Lagrangian reads
\begin{eqnarray}
L_{\rm int} =  F(g_{ij},R_{kli}{}^j,T_{kl}{}^i) L_{\rm mat},\label{ansatz_lagrangian_model_3}
\end{eqnarray}
and allow for a coupling function $F(g_{ij},R_{kli}{}^j,T_{kl}{}^i)$ to be a function of independent scalar invariants constructed in all possible ways from the components of the curvature and torsion tensors.  The matter Lagrangian has the usual form $L_{\rm mat} = L_{\rm mat}(\psi^A, \nabla_i\psi^A, g_{ij})$. A Lagrange-Noether analysis, see Ref.~\refcite{Obukhov:Puetzfeld:2013}, yields the general conservations laws of the theory, i.e.\
\begin{eqnarray}
\widehat{\nabla}_n \tau_{[ik]}{}^n &=& K_{ni}{}^l \tau_{[kl]}{}^n - K_{nk}{}^l \tau_{[il]}{}^n - \Sigma_{[ik]} - A_n \tau_{[ik]}{}^n, \label{re_cons1c}\\
\widehat{\nabla}_i \Sigma_k{}^i &=& - \Sigma_l{}^i K_{ki}{}^l - \tau^m{}_n{}^l R_{klm}{}^n - A^i \Xi_{ik}- A_i \Sigma_k{}^i. \label{recons2c}
\end{eqnarray}
Here 
\begin{equation}
\Sigma_k{}^i = {\frac {\partial {L_{\rm mat}}}{\partial\nabla_i\psi^A}} \,\nabla_k\psi^A - \delta^i_kL_{\rm mat},\label{emcan}
\end{equation}
denotes the canonical energy-momentum tensor, and
\begin{equation}
\tau^n{}_k{}^i = -\,{\frac {\partial {L_{\rm mat}}}{\partial\nabla_i\psi^A}} \,(\sigma^A{}_B)_k{}^n \psi^B,\label{spin}
\end{equation}
the canonical spin tensor. Furthermore, we made use of the shortcut $\Xi_{ij}:=g_{ij}L_{\rm mat}$, and auxiliary variables like in Ref.~\refcite{Puetzfeld:Obukhov:2013}, i.e.\ $A(g_{ij},R_{ijk}{}^l,T_{ij}{}^k):=\log F$, $A_i:=\nabla_i A$, $A_{ij}:=\widehat{\nabla}_j \nabla_i A$ etc. The Riemann-Cartan connection was decomposed into the Riemannian (Christoffel) connection\footnote{We use the hat to denote objects and operators (such as the curvature, covariant derivatives, etc) defined by the Riemannian connection (\ref{chr}).}
\begin{equation}
\widehat{\Gamma}_{ij}{}^k = \left\{{}_{ij}^k\right\} = {\frac 12}g^{kl}\left(\partial_ig_{jl} + \partial_jg_{il} - \partial_lg_{ij}\right),\label{chr}
\end{equation} 
plus the post-Riemannian piece: 
\begin{equation}
{\Gamma}_{ij}{}^k = \widehat{\Gamma}_{ij}{}^k - K_{ij}{}^k.\label{GG}
\end{equation} 
Here the contortion tensor reads
\begin{equation}\label{RCconn}
K_{ij}{}^k = -\,{\frac 12}(T_{ij}{}^k - T_{j}{}^k{}_i + T^k{}_{ij}) = - K_{i}{}^k{}_j .
\end{equation}

\section{Equations of motion}\label{eom_sec}

The conservation equations (\ref{re_cons1c}) and (\ref{recons2c}) form the basis for a general multipolar analysis. Utilizing the geodesic expansion technique of Synge \cite{Synge:1960}, one can derive equations of motion of a test body \cite{Puetzfeld:Obukhov:2013:2}. For this we use the world-function $\sigma$ and the parallel propagator $g^{y}{}_{x}$ and denote 
\begin{eqnarray}
\Phi^{y_1\dots y_ny_0}{}_{x_0} &:=& \sigma^{y_1} \cdots \sigma^{y_n} g^{y_0}{}_{x_0},\label{Phi}\\
\Psi^{y_1\dots y_ny_0y'}{}_{x_0x'} &:=& \sigma^{y_1} \cdots \sigma^{y_n} g^{y_0}{}_{x_0}g^{y'}{}_{x'}.
\label{Psi}
\end{eqnarray}
Furthermore, we introduce integrated moments \`{a} la Dixon \cite{Dixon:1964} of an arbitrary order $n=0,1,2,\dots$:
\begin{eqnarray}
p^{y_1 \dots y_n y_0}&:=& (-1)^n  \int\limits_{\Sigma(s)}\Phi^{y_1\dots y_ny_0}{}_{x_0} \widetilde{\Sigma}^{x_0 x_1} d \Sigma_{x_1}, \label{p_moments_def} \\
t^{y_2 \dots y_{n+1} y_0 y_1}&:=& (-1)^{n}  \int\limits_{\Sigma(s)} \Psi^{y_2\dots y_{n+1}y_0y_1}{}_{x_0x_1} \widetilde{\Sigma}^{x_0 x_1} w^{x_2} d \Sigma_{x_2}, \label{t_moments_def} \\
\xi^{y_2 \dots y_{n+1} y_0 y_1}&:=& (-1)^{n}  \int\limits_{\Sigma(s)} \Psi^{y_2\dots y_{n+1}y_0y_1}{}_{x_0x_1} \widetilde{\Xi}^{x_0 x_1} w^{x_2} d \Sigma_{x_2}, \label{xi_moments_def} \\
s^{y_2 \dots y_{n+1} y_0 y_1}&:=& (-1)^{n}  \int\limits_{\Sigma(s)} \Psi^{y_2\dots y_{n+1}y_0y_1}{}_{x_0x_1}  \widetilde{\tau}^{[x_0 x_1] x_2} d \Sigma_{x_2}, \label{s_moments_def}\\
q^{y_3 \dots y_{n+2} y_0 y_1 y_2}&:=& (-1)^{n}  \int\limits_{\Sigma(s)} \Psi^{y_3\dots y_{n+2}y_0y_1}{}_{x_0x_1} g^{y_2}{}_{x_2}  \widetilde{\tau}^{[x_0 x_1] x_2} w^{x_3} d \Sigma_{x_3}. \label{q_moments_def}
\end{eqnarray}
Here the integrals are performed over spatial hypersurfaces. Note that in our notation the point to which the index of a bitensor belongs can be directly read from the index itself; e.g., $y_{n}$ denotes indices at the point $y$. Furthermore, we will now associate the point $y$ with the world-line of the test body under consideration.

\section{Nonminimal coupling}\label{nonminimal_subsubsec}

A general extended body consists of material elements with microstructure, i.e., with spin. In the pole-dipole approximation, the relevant moments are $p^a, p^{ab}, t^{ab}, t^{abc}, \xi^{ab}, \xi^{abc}, s^{ab}, q^{abc}$. If we neglect all higher multipole moments and introduce the {\it integrated orbital angular momentum} and the {\it integrated spin angular momentum} of an extended body as
\begin{equation}
L^{ab} := 2p^{[ab]},\qquad S^{ab} := -\,2s^{ab},\label{LS}
\end{equation}
we obtain the following equations of motion:
\begin{eqnarray}
{\frac{D}{ds}}{\cal J}^{ab} &=& -\,2v^{[a}{\cal P}^{b]} + 2FQ^{cd[a}T_{cd}{}^{b]} + 4FQ^{[a}{}_{cd}T^{b]cd}\nonumber\\
&& -\,\left(4q^{[a|c|b]} + 2\xi^{[a|c|b]}\right)\nabla_cF,\label{eq_rot}\\
{\frac{D}{ds}}{\cal P}^{a} &=& {\frac 12}\widehat{R}^a{}_{bcd}{\cal J}^{cd}v^b + FQ^{bc}{}_d\widehat{\nabla}{}^a T_{bc}{}^d\nonumber\\
&& -\,2q^{bcd}K_{dc}{}^a\nabla_bF + 2Fq^{acd}\nabla_dA_c\nonumber\\ 
&& -\,\xi^{ba}\nabla_bF - \xi^{cba}\widehat{\nabla}_c\nabla_bF.\label{eq_transl}
\end{eqnarray}
Here $v^{y}:=dx^{y}/ds$, $s$ is the proper time, ${\frac{D}{ds}} = v^i\widehat{\nabla}_i$, and we defined the total energy-momentum vector and the total angular momentum tensor by
\begin{eqnarray}\label{Pa}
{\cal P}^a &:=& F\left(p^a - {\frac 12}K^a{}_{cd}S^{cd}\right) + \left(p^{ba} - S^{ab}\right)\nabla_bF,\\
{\cal J}^{ab} &:=& F\left(L^{ab} + S^{ab}\right).\label{Jab}
\end{eqnarray}
In addition, we introduced a redefined moment
\begin{equation}
Q^{bca} := {\frac 12}\left(q^{bca} + q^{bac} - q^{cab}\right).\label{Qabc}
\end{equation}

The equations of motion (\ref{eq_rot}) and (\ref{eq_transl}) generalize the results obtained in Ref.~\refcite{Puetzfeld:Obukhov:2013} to the case when extended bodies are built of matter with microstructure and move in a Riemann-Cartan spacetime with nontrivial torsion. 

\section{Minimal coupling}\label{minimimal_dipole_subsubsec}

When the coupling function is constant, $F = 1$, that is for the {\it minimal coupling} case, we obtain 
\begin{equation}\label{noF}
{\cal P}^a = p^a - {\frac 12}K^a{}_{cd}S^{cd},\qquad {\cal J}^{ab} = L^{ab} + S^{ab},
\end{equation}
and the equations of motion 
\begin{eqnarray}
{\frac{D}{ds}}{\cal J}^{ab} &=& -\,2v^{[a}{\cal P}^{b]} + 2Q^{cd[a}T_{cd}{}^{b]} + 4Q^{[a}{}_{cd}T^{b]cd},\label{eq_rot_noF}\\
{\frac{D}{ds}}{\cal P}^{a} &=& {\frac 12}\widehat{R}^a{}_{bcd}{\cal J}^{cd}v^b + Q^{bc}{}_d\widehat{\nabla}{}^a T_{bc}{}^d.\label{eq_transl_noF}
\end{eqnarray}
It is satisfying to see that the structure of the equations of motion for minimal coupling  (\ref{eq_rot_noF})-(\ref{eq_transl_noF}) is in agreement with the earlier results of Yasskin and Stoeger \cite{Stoeger:Yasskin:1980}. Therefore, we confirm once again that spacetime torsion {\it in the minimal coupling scheme} interacts only with the integrated spin $S^{ab}$, which arises from the intrinsic spin of matter, and the higher moment $q^{abc}$. Hence, usual matter without microstructure cannot detect torsion and, in particular, experiments with macroscopically rotating bodies such as gyroscopes in the Gravity Probe B mission do not place any limits on torsion \cite{Hehl:Obukhov:Puetzfeld:2013}. 

\section{Nonminimal coupling: a loophole to detect torsion?}\label{nonminimimal_dipole_subsubsec}

However, the conclusion that matter without microstructure cannot detect torsion is apparently violated for the {\it nonminimal coupling} case. As we see from (\ref{eq_rot}) and (\ref{eq_transl}), test bodies of structureless matter could be affected by torsion via the derivatives of the coupling function $F(g_{ij},R_{kli}{}^j,T_{kl}{}^i)$. This possibility, however, is qualitatively different from the ad hoc assumption that structureless particles move along auto-parallel curves in the Riemann-Cartan spacetime made in Refs.~\citen{Kleinert:1998,Mao:2009,March:2011a,March:2011b}; see the critical assessment in Ref.~\refcite{Hehl:Obukhov:Puetzfeld:2013}. The trajectory of a monopole particle without intrinsic spin ($\tau_{ab}{}^c=0$), is described by 
\begin{equation}
{\frac{D}{ds}} (Fp^{a}) = -\,\xi^{ab}\nabla_bF,\label{mono_eom_3}
\end{equation}
which is neither geodesic nor auto-parallel. The same is true for the dipole case when the nonminimal coupling force is combined with the Mathisson-Papapetrou force.

\section{Conclusions}\label{conclusion_sec}

We have presented equations of motion for material bodies with microstructure for a very large class of gravitational theories, thus extending the previous works \cite{Bailey:Israel:1975,Stoeger:Yasskin:1979,Stoeger:Yasskin:1980,Puetzfeld:Obukhov:2007,Puetzfeld:Obukhov:2008:1} to the general framework with nonminimal coupling. In the special case of minimal coupling (which is recovered when $F=1$), our results can be viewed as the covariant generalization of the ones in Refs.~\citen{Stoeger:Yasskin:1979,Stoeger:Yasskin:1980}, as well as the parts concerning Poincar\'{e} gauge theory of Ref.~\refcite{Puetzfeld:Obukhov:2007}. 

A somewhat surprising result in the present nonminimal context with torsion, is the -- indirect -- influence of the torsion through the coupling function $F$ on the dynamics of matter without intrinsic spin even in the lowest order equations of motion -- see eq.\ (\ref{mono_eom_3}). This clearly is a distinctive feature of theories which exhibit nonminimal coupling, which sets them apart from other gauge theoretical approaches to gravity. 

Experiments testing the universality of free fall should be used to put strong limits on theories  with nonminimal coupling (already present day accelerometers reach a sensitivity of $<10^{-12}$ m/s$^2$). Experimentalists are thus encouraged to use our results as a universal framework to systematically test the effects of nonminimal coupling by means of spinning, as well as structureless massive test bodies.

\section*{Acknowledgements}
This work was supported by the Deutsche Forschungsgemeinschaft (DFG) through the grant LA-905/8-1/2 (D.P.).
 
\bibliographystyle{ws-ijmpd}

\bibliography{puetzfeld_obukhov_essay_2014}

\begin{thebibliography}{10}

\bibitem{Einstein:1921}
A.~{Einstein}, {\em Sitzungsber. preuss. Akad. Wiss.} {\bf 1}  (1921)   123.

\bibitem{Hehl:1995}
F.~W. {Hehl}, J.~D. {McCrea}, E.~W. {Mielke} and Y.~{Ne'eman}, {\em Phys. Rep.}
  {\bf 258}  (1995)  ~1.

\bibitem{Blagojevic:Hehl:2013}
M.~{Blagojevi\'c} and F.~{Hehl}, {\em {Gauge Theories of Gravitation. A Reader
  with Commentaries}} (Imperial College Press, London, 2013).

\bibitem{Stoeger:Yasskin:1980}
P.~B. {Yasskin} and W.~R. {Stoeger}, {\em Phys. Rev. D} {\bf 21}  (1980)
  2081.

\bibitem{Hunter:etal:2013}
L.~{Hunter}, J.~{Gordon}, S.~{Peck}, D.~{Ang} and J.-F. {Lin}, {\em Science}
  {\bf 339}  (2013)   928.

\bibitem{Lehnert:etal:2014}
R.~{Lehnert}, W.~M. {Snow} and H.~{Yan}, {\em Phys. Lett. B} {\bf 730}  (2014)
   353.

\bibitem{Obukhov:Puetzfeld:2013}
Y.~N. {Obukhov} and D.~{Puetzfeld}, {\em Phys. Rev. D} {\bf 87}  (2013)
  081502(R).

\bibitem{Puetzfeld:Obukhov:2013}
D.~{Puetzfeld} and Y.~N. {Obukhov}, {\em Phys. Rev. D} {\bf 87}  (2013)
  044045.

\bibitem{Synge:1960}
J.~L. {Synge}, {\em {Relativity: The general theory}} (North-Holland,
  Amsterdam, 1960).

\bibitem{Puetzfeld:Obukhov:2013:2}
D.~{Puetzfeld} and Y.~N. {Obukhov}, {\em Phys. Rev. D} {\bf 88}  (2013)
  064025.

\bibitem{Dixon:1964}
W.~G. {Dixon}, {\em Nuovo Cimento} {\bf 34}  (1964)   317.

\bibitem{Hehl:Obukhov:Puetzfeld:2013}
F.~W. {Hehl}, Y.~N. {Obukhov} and D.~{Puetzfeld}, {\em Phys. Lett. A} {\bf 377}
   (2013)   1775.

\bibitem{Kleinert:1998}
H.~Kleinert, {\em Gen. Rel. Grav.} {\bf 32}  (2000)   1271.

\bibitem{Mao:2009}
Y.~{Mao}, M.~{Tegmark}, A.~{Guth} and S.~{Cabi}, {\em Phys. Rev. D} {\bf 76}
  (2007)   104029.

\bibitem{March:2011a}
R.~{March}, G.~{Bellettini}, R.~{Tauraso} and S.~{Dell'Agnello}, {\em Phys.
  Rev. D} {\bf 83}  (2011)   104008.

\bibitem{March:2011b}
R.~{March}, G.~{Bellettini}, R.~{Tauraso} and S.~{Dell'Agnello}, {\em Gen. Rel.
  Grav.} {\bf 43}  (2011)   3099.

\bibitem{Bailey:Israel:1975}
I.~{Bailey} and W.~{Israel}, {\em Comm. Math. Phys.} {\bf 42}  (1975)  ~65.

\bibitem{Stoeger:Yasskin:1979}
W.~R. {Stoeger} and P.~B. {Yasskin}, {\em Gen. Rel. Grav.} {\bf 11}  (1979)
  427.

\bibitem{Puetzfeld:Obukhov:2007}
D.~{Puetzfeld} and Y.~N. {Obukhov}, {\em Phys. Rev. D} {\bf 76}  (2007)
  084025.

\bibitem{Puetzfeld:Obukhov:2008:1}
D.~{Puetzfeld} and Y.~N. {Obukhov}, {\em Phys. Rev. D} {\bf 78}  (2008)
  121501.

\end{thebibliography}

\end{document}